\documentclass[10pt]{article}

 \usepackage{amsmath,amsthm,amssymb}
 \usepackage{makeidx,epsfig,lscape}
 \usepackage{color,colortbl}
 \usepackage{fancyhdr}
 \usepackage{xcolor,pict2e}

 \definecolor{myaqua}{rgb}{0.0,0.5,0.55}
 \definecolor{lightaqua}{rgb}{0.75,0.95,0.95}

 \usepackage[colorlinks = true,
            linkcolor = myaqua,
            urlcolor  = blue,
            citecolor = myaqua]{hyperref}

\usepackage{caption}
\usepackage{floatrow}
\def\bt{\begin{tabular}}
\def\et{\end{tabular}}
\def\and{\mbox{ and }}

\def\1{{\bf 1}}

\begin{document}
\DeclareGraphicsRule{.bmp}{bmp}{}{}
\newcommand{\cao}{\c{c}\~ao}
\newcommand{\coes}{\c{c}\~oes}
\newcommand{\caO}{\c{c}\c}
\newcommand{\coeS}{\c{c}\~oes}
\newcommand{\CAO}{\c{C}\~AO}
\newcommand{\COES}{\c{C}\~OES}
\newcommand{\cc }{\c{c}}
\newcommand{\CC }{\c{C}}
\newcommand{\rg}{\sqrt{g} }
\def\sen{\rm{sen\,}}
\def\ph#1{\phi_{#1}}
\def\br#1{\stackrel{-}#1}
\def\n{\noindent}
\def\ep{\epsilon}

 \vskip 12mm

{ 

{\noindent{\huge\bf\color{myaqua}
  Performance Measure for Optimal Quantum Control}}

{\large\bf Alexandre Coutinho Lisboa$^1$, Jos\'e Roberto Castilho Piqueira$^2$}}
\\[2mm]
{ 
 $^1$Department of Telecommunication and Control Engineering, Universidade de S\~ao Paulo, S\~ao Paulo-Brazil.\\
Email: \href{mailto:alexandrecl@usp.br}{\color{blue}{\underline{\smash{alexandrecl@usp.br}}}}\\[1mm]
$^2$Department of Telecommunication and Control Engineering,
Universidade de S\~ao Paulo, S\~ao Paulo-Brazil.
Email:
\href{mailto:piqueira@lac.usp.br}{\color{blue}{\underline{\smash{piqueira@lac.usp.br}}}}
\\[4mm]
Received **** 2015
 \\[4mm]

 { 
{\noindent{\large\bf\color{myaqua} Abstract}{\bf \\[3mm]
\textup{
The problem concerning the minimum time for an initial state to evolve up to a target state plays an important role in the
Classic Optimal Control theory. In the quantum context, the problem is more complex since quantum states are highly sensitive to environmental influences. However, its formulation is decisive for building quantum information processing systems. Quantum evolution time has been studied from the theoretical point, being known as the Quantum Speed Limit (QSL) problem, providing general results.
Considering the implementation of quantum control systems, as the decoherence phenomenon is unavoidable, it is important to apply these general
results to particular cases, developing expressions and performance
measures, in order to assist control engineering designers. Here, the time-energy uncertainty relations are revisited, being fundamental for proposing performance measures based on minimum time evolution. A minimum time performance measure is defined
for quantum control problems and applied to some practical examples, providing practical hints that are supposed to be useful for researchers pursuing optimization strategies for quantum control
systems.
 }}}

{\noindent{\large\bf\color{myaqua} Keywords}{\bf \\[3mm]
 Minimum Time; Optimal Control; Quantum Systems; Time-Energy Uncertainty.
}

\section{Introduction}
Throughout the $20^{th}$ century, Physics and Cybernetics experienced a strong development decisively contributing to modern science regardless of the weak
interactions between them. This, can be due to the different approaches with Physics being a descriptive science and Cybernetics a prescriptive one \cite{1}.

However, it can not be denied that automatic systems play an important role concerning physical experiments, but in most cases the control theory is considered
to be secondary with no effective contributions for explaining physical phenomena \cite{1}.

In the late 1980s, with the development of ultra fast lasers, methods based on optimal control were developed to control molecular systems \cite{2}. The interest in this
kind of problem increased in the early 1990s and the concepts of classical and quantum approaches were developed \cite{3,4}.

Additionally, researches on quantum computation hardware can be implemented by manipulating the quantum state of trapped ions via laser or electrical fields \cite{5}.

Consequently, it seems to be important to formulate optimal control strategies for quantum particles allowing, for example, to minimize times, average
distances or energy costs of the processes.

Luo \cite{5a} developed a theoretical method to calculate the minimum time for a state evolution, followed by Chau \cite{6c3}, and applied to determine the quantum speed limit (QSL) \cite{51,52}. The work conducted here is dedicated to analyze how the time-energy uncertainty affects the expression of the minimum time for state evolution and proposing performance measures for this kind of process, aiming at defining some practical hints for quantum control designers.

This paper starts with a section explaining the derivation of time-energy uncertainty relation, following the seminal work by  Mandelstam and Tamm \cite{5b}, resulting an expression to the minimum time between states, called Bhattacharyya limit \cite{5c}, which is valid considering the constant Hamiltonian evolution.

In the next section, the Bhattacharyya limit is used to define a performance measure, essential to design quantum control devices, allowing an evaluation criterion to control systems from the minimum time optimization point of view \cite{6}, with quantum fidelity playing an important role on defining the reacheability of the target state.

Three essential problems to quantum control design are explored in a new section, in order to show the practical application of the defined performance measure. The first one regards to the transition between two orthogonal states, essential for quantum information processing. The second is related to the Fahri and Gutmann formulation \cite{6a} for digital quantum computation. The third considers a general maximum fidelity \cite{6b} state transition. A conclusion section finishes the work.

\section{Mandelstam-Tamm uncertainty relations and Bhattacharyya limit}

The first theoretically satisfactory formulation of the time-energy uncertainty was proposed by Leonid Mandelstam and Igor Tamm in 1945 \cite{5b}. Fock and Krylov obtained the same results in 1947 \cite{6c1}, and the problem was theoretically treated by several research, providing interesting variants of the original work for particular situations, summarized by Dodonov and Dodonov \cite{6c2}.

In this section, the time-energy uncertainty is derived in a slightly different form, for simplifying the expressions of the performance measure to be defined.

Considering the quantum observables $\hat{R}$ and $\hat{S}$, the following relations can be written \cite{5b}:

\begin{equation}
\Delta\hat{S}.\Delta\hat{R} \geq \frac{1}{2}|<\hat{R}\hat{S}-\hat{S}\hat{R}>|;
\label{eq1}
\end{equation}
\noindent and

\begin{equation}
\frac{d<\hat{R}>}{dt} = \frac{1}{i\hbar}<[\hat{R},\hat{H}]>.
\label{eq2}
\end{equation}

In expression (\ref{eq1}), $\Delta\hat{S}$ and $\Delta\hat{R}$ are the standard deviations of operators $\hat{S}$ and $\hat{R}$, respectively. Equation (\ref{eq2}) is the dynamical evolution for the mean value of operator $\hat{R}$.

By making $\hat{S}=\hat{H}$ and considering the relations (\ref{eq1}) and (\ref{eq2}), one can write:

\begin{equation}
\Delta\hat{H}.\Delta\hat{R} \geq \frac{\hbar}{2} |\frac{d<\hat{R}>}{dt}|;
\label{eq3}
\end{equation}
\noindent giving the relation between the total energy standard deviation $\Delta\hat{H}$, or energy uncertainty, and the uncertainty of another dynamical quantity, relating them with the mean value of this quantity.

As the absolute value of the integral of a function is lower than or equal to the integral of the absolute value of the function, (\ref{eq3}) can be integrated between times $t$ and $t+\Delta t$, with $H$ constant during this interval, resulting:

\begin{equation}
\Delta\hat{H}.\Delta t \geq \frac{\hbar}{2}\frac{|<R_{t+\Delta t}>-<R_{t}>|}{<\Delta\hat{R}>};
\label{eq4}
\end{equation}

\noindent with $<\Delta\hat{R}>$ being the mean value of $\Delta\hat{R}$ in the time interval $\Delta t$.

From now on, the  $\Delta t$ represents the minimum time for the mean value of a physical quantity to be varied by its standard deviation. Consequently, $\Delta t$ is the uncertainty in time and (\ref{eq4}) becomes:

\begin{equation}
\Delta\hat{H}.\Delta t \geq \frac{\hbar}{2}.
\label{eq5}
\end{equation}

Considering the projection operator $\hat{\Lambda} = |\Psi_0><\Psi_0|$, with $\hat{\Lambda}^{2} = \hat{\Lambda}\hat{\Lambda} = \hat{\Lambda}$, its mean value can be seen as the probability of the system to be in a quantum state $|\Psi>$. This fact can be justified as:

\begin{equation}
\hat{\Lambda}_{\Psi} = <\Psi|\hat{\Lambda}|\Psi> = <\Psi|\Psi_0><\Psi_0|\Psi> = |<\Psi|\Psi_0>|^{2} = P_{\Psi}.
\label{eq6}
\end{equation}

It can be noticed that $0 \leq <\hat{\Lambda}> \leq 1 $ and, considering the definition of the
standard deviation:

\begin{equation}
\Delta\hat{\Lambda} = \sqrt {<\hat{\Lambda}^{2}> - <\hat{\Lambda}>^{2}} = \sqrt {<\hat{\Lambda}> - <\hat{\Lambda}>^{2}}.
\label{eq7}
\end{equation}

Replacing $\hat{R}$ by $\hat{\Lambda}$ in (\ref{eq3}) and considering expression (\ref{eq7}), results:

\begin{equation}
\Delta\hat{H}.\sqrt {<\hat{\Lambda}> - <\hat{\Lambda}>^{2}} \geq \frac{\hbar}{2} \frac{d<\hat{\Lambda}>}{dt}.
\label{eq8}
\end{equation}

Expression (\ref{eq8}) contains only the quantities $\hat{\Lambda} = \hat{\Lambda}(t)$ and its derivative depending on time. Consequently, it can be integrated in time. If, for instance, $\hat{\Lambda}(0) = 1$, indicating that for $t=0$ it is certain that the state is $\Psi_0$, for $t \geq 0$, the integration gives:

\begin {equation}
\frac{\pi}{2} - \arcsin \sqrt{<\hat{\Lambda}(t)>} \leq \frac {\Delta\hat{H}.t}{\hbar}.
\label{eq9}
\end{equation}

By using some trigonometric identities and algebraic manipulation, (\ref{eq9}) is transformed into:
\begin{equation}
<\hat{\Lambda}(t)> \, \geq \cos^{2}(\frac {\Delta\hat{H}.t}{\hbar}).
\label{eq10}
\end{equation}

As the mean value of the operator $\hat{\Lambda}$ corresponds to the probability of finding the system in the state $<\Psi_t>$ at time t, starting from state $<\Psi_0>$ at $t=0$, that is denoted by $P_t$, expression (\ref{eq10}) is rewritten as:

\begin{equation}
P_t \geq \cos^{2}(\frac {\Delta\hat{H}.t}{\hbar}).
\label{eq11}
\end{equation}

Observing (\ref{eq11}), the time physically possible for a state transition, considering a time independent Hamiltonian, is given by:

\begin{equation}
t \geq \frac{\hbar}{\Delta\hat{H}} \arccos \sqrt {P_t},
\label {eq12}
\end{equation}
\noindent considering the standard deviation $\Delta\hat{H}$, t the transition time and $P_t = |<\Psi_t|\Psi_0>|^{2}$

Expression (\ref{eq12}) is normally called ``Bhattacharyya limit" due to the important work presented  by K. Bhattacharyya in 1983\cite{5c}.

\section{Minimum Time Performance Measure}

The ideas developed by Luo \cite{5a} were improved in several theoretical studies about the speed limit for quantum systems \cite{6c3} and its relation with
entanglement.

Here, the intention is to connect this theoretical point of view to problems concerning practical quantum control systems, aiming at defining a quantity for measuring how far from the optimal conditions the operational conditions are, in order give tools to quantum control design.

Consequently, a minimum time performance measure is proposed, $\eta_t$ defined as:

\begin{equation}
\eta_t = \frac{t_{min}}{t_{CQS}},
\label{eq13}
\end{equation}
\noindent with $t_{min}$ representing the inferior limit of expression (\ref{eq12}) and $t_{CQS}$, the effective time for the target transition state.

The minimum time performance measure $\eta_t$ is a real number belonging to the interval $[0,1]$. If the state transition has not occured or the control algorithm has not converged, it is considered that $t_{CQS} \rightarrow \infty$, i.e., $\eta_t =0$. For the ideal state transition $(t_{QCS}=t_{min})$, $\eta_t = 1$.

Taking expression $(\ref{eq12})$ into the minimum time performance measure, its complete expression becomes:

\begin{equation}
\eta_t = \frac{\hbar \arccos \sqrt {P_t}}{\Delta\hat{H}.t_{CQS}}.
\label {eq14}
\end{equation}

If a specific state transition between an initial state $|\Psi_I>$ and a target state $|\Psi_G>$ is considered, expression (\ref{eq14}) is modified to:

\begin{equation}
\eta_t = \frac{\hbar \arccos |<\Psi_G|\Psi_I>|}{\Delta\hat{H}.t_{CQS}}.
\label {eq15}
\end{equation}

Frequently, it is not possible for practical applications or numerical simulations of optimal control algorithms to obtain the exact transition for the target state. In these cases, it is important to find optimal controls $u^{*}(t)$ in order to maximize de quantum fidelity $F$ \cite{6b,6c4} between the final state $|\Psi_F>$ and the target state $<\Psi_G>$, defined as:

\begin{equation}
F = |<\Psi_G|\Psi_F>|^{2}.
\label{eq16}
\end{equation}

\section{Performance measure for particular state transitions}

In this section, three different state transitions are considered and, in each case, the minimum time performance measure is calculated.

\subsection{Transition between two orthogonal states}

Considering the transition between an initial state $\Psi$ and its orthogonal state $\Psi^{\perp}$, occurring in a time $t$, as $P_t = |<\Psi^{\perp}|\Psi>|^{2}$, equation (\ref{eq11}) gives:

\begin{equation}
|<\Psi^{\perp}|\Psi>|^{2} \geq \cos^{2}(\frac {\Delta\hat{H}.t}{\hbar}).
\label{eq17}
\end{equation}

The dynamical evolution between these states is subjected to a temporal evolution operator $\hat{U}(t,t_0)$, with $\hat{U}(t,t_0) = e^{-{\frac{i\hat{H}(t-t_0)}{\hbar}}}$ \cite{6c}, for a time independent Hamiltonian $\hat{H}$.

Consequently:

\begin{equation}
|<\Psi|\hat{U}(t,t_0)|\Psi>|^{2} \geq \cos^{2}(\frac {\Delta\hat{H}.t}{\hbar}),
\label{eq18}
\end{equation}
\noindent with $|<\Psi|\hat{U}(t,t_0)|\Psi>|^{2}$ considered to be the surviving probability of state $\Psi$, while $\hat{U}(t,t_0)$ acts over the system.

If the system evolves from state $\Psi$ to its orthogonal state $\Psi^{\perp}$, this probability vanishes, i.e., the transition time obeys:

\begin{equation}
t_{\Psi \rightarrow \Psi^{\perp}} = \inf \{{t \geq 0: <\Psi|\hat{U}(t,t_0)|\Psi> = 0}\},
\label{eq19}
\end{equation}
\noindent and, therefore:

\begin{equation}
t_{\Psi \rightarrow \Psi^{\perp}} \geq \frac{\pi\hbar}{2\Delta\hat{H}}.
\label{eq20}
\end{equation}

Hence, the minimum time performance measure defined by (\ref{eq13}) for transitions between orthogonal states is given by:

\begin{equation}
\eta_{\Psi \rightarrow \Psi^{\perp}}^{\hat{H}} = \frac{\pi\hbar}{2\Delta\hat{H}t_{CQS}}.
\label{eq21}
\end{equation}

\subsection{Digital quantum computation model}

Here an example related to digital quantum computation \cite{6a}, which can be formulated as a quantum searching algorithm \cite{6d}, is studied in order to derive its minimum time performance measure.

The initial idea is to consider $|a>$ and $|b>$ as initial and target states, respectively with the system Hamiltonian given by:

\begin{equation}
\hat{H} = E_ a |a><a| + E_b |b><b|,
\label{eq22}
\end{equation}
\noindent with $E_a$ and $E_b$ positive constants.

As the exact formulation for the evolution from state $|a>$ to $|b>$ is almost impossible, an alternative formulation that considers fidelity $F$ is proposed. The reasoning is to obtain the minimum possible time, in order to maximize $F$, given by:

\begin{equation}
F = P_t = |<b|\hat{U}(t,t_0)|a>|^{2}.
\label{eq23}
\end{equation}

In order to perform the calculations, it is necessary to choose a normalized orthogonal basis in the space generated by $|a>$ and $|b>$, composed of the kets $|b>$ and $|b^{'}> = \frac{1}{\sqrt{1-s^{2}}}(|a> - s|b>)$, with $s=<a|b>$.

On this basis, $|a>$ and $|b>$ are expressed as:

\begin{eqnarray}
|a>=
 \left [
   \begin{array}{c}
   s  \\
   \sqrt{1-s^{2}} \\
   \end{array}
 \right],
 |b>=
 \left [
   \begin{array}{c}
   1  \\
   0 \\
   \end{array}
 \right].
\end{eqnarray}

Considering these expressions of the kets $|a>$ and $|b>$ on the new basis, the Hamiltonian becomes:

\begin{eqnarray}
\hat{H}=
 \left [
   \begin{array}{cc}
   1+s^{2}-\frac{x}{E}(1-s^{2}) & (1+\frac{x}{E})s\sqrt{1-s^{2}}  \\
   (1+\frac{x}{E})s\sqrt{1-s^{2}} & 1-s^{2}+\frac{x}{E}(1-s^{2}) \\
   \end{array}
 \right]
 =\frac{E}{2}
 \left [
   \begin{array}{cc}
   1+\lambda & \sqrt{\mu^{2}-\lambda^{2}}  \\
   \sqrt{\mu^{2}-\lambda^{2}} & 1-\lambda\\
   \end{array}
 \right],
\end{eqnarray}
\noindent with $E=E_a + E_b$, $x=E_a - E_b$, $\mu=\sqrt{s^{2}+(\frac{x}{E})^{2}(1-s^{2})}$ and $\lambda=s^{2}-(\frac{x}{E})(1-s^{2})$.

From these considerations, it is possible to diagonalize the Hamiltonian operator as follows:

\begin{eqnarray}
\hat{H}= U
 \left [
   \begin{array}{cc}
   \frac{E}{2}(1+\mu) & 0  \\
   0 & \frac{E}{2}(1-\mu) \\
   \end{array}
 \right]U^{-1},
\end{eqnarray}
\noindent with $U$ given by:
\begin{eqnarray}
U = \frac{1}{\sqrt{2}}
 \left [
   \begin{array}{cc}
   \sqrt{1 + \frac{\lambda}{\mu}} & \sqrt{1 - \frac{\lambda}{\mu}}  \\
   \sqrt{1 - \frac{\lambda}{\mu}}& -\sqrt{1 + \frac{\lambda}{\mu}} \\
   \end{array}
 \right]U^{-1},
\end{eqnarray}
\noindent corresponding to the unitary operator constructed with the eigenvalues of $\hat{H}$.

Taking into account equations (22) to (27), it is possible to obtain the fidelity, expressed by the probability of, starting with state $|a>$, following the dynamical operator $\hat{U}(t,t_0) = e^{-{\frac{i\hat{H}(t-t_0)}{\hbar}}}$, to reach the target state $|b>$. i.e.:

\begin{equation}
P_t = |<b|e^{-{\frac{i\hat{H}t}{\hbar}}}|a>|^{2} = s^{2}[(\frac{1}{\mu^{2}}-1)\sin^{2}(\frac{\mu
Et}{2\hbar})+1],
\label{eq30}
\end{equation}
\noindent considering $t_0=0$.

It can be noticed that $s \leq \mu \leq 1$; therefore, maximum value of the probability $P_t$, imposing $\sin^{2}(\frac{\mu
Et}{2\hbar})=1$, is:

\begin{equation}
P_{max} = max P_t (t \geq 0) = (\frac{s}{\mu})^{2}.
\label{eq31}
\end{equation}

Consequently, it is natural to derive the expression for $t_{min}$, the minimum time to obtain maximum fidelity, by simply applying the same conditions above. Therefore:

\begin{equation}
t_{min} = inf\{{t \geq 0: P_t = P_{max}}\} = \frac{\pi\hbar}{E\mu}.
\label{eq31}
\end{equation}

Replacing equation (\ref{eq31}) in definition (\ref{eq15}), the expression for measuring minimum time optimal control performance in a Fahri-Gutmann system \cite{6a}, with Hamiltonian independent of time, is:

\begin{equation}
\eta_{FG}^{H} = \frac{\pi\hbar}{E\mu t_{CQS}}.
\label{eq32}
\end{equation}

\subsection{General state transition}

Here, the minimum time quantum control performance measure will be derived for a transition between two general states. In order to obtain this expression, it is necessary to start with a transition between a state and the auxiliary state orthogonal to the other state, deriving an intermediate expression that is used to complete the task.

\subsubsection{Transition between a state and a state orthogonal to an other}

Considering the states $|a>$ and $|c>$, with $|<c|a>|^{2} = \cos \phi$, and $\phi \in [0, \frac{\pi}{2}]$, if the dynamical evolution of the system is subjected to a temporal evolution operator $\hat{U}(t,t_0)$, with $\hat{U}(t,t_0) = e^{-{\frac{i\hat{H}(t-t_0)}{\hbar}}}$, for a time independent Hamiltonian $\hat{H}$, it is possible to study the state transition between $|a>$ and $|c^{\perp}>$, orthogonal to $|c>$, by the temporal transition probability:

\begin{equation}
F = P_t = |<c|\hat{U}(t,t_0)|a>|^{2}.
\label{eq33}
\end{equation}

Defining $\hat{R} = |c><c|$, $\Delta\hat{R}=\sqrt{P_t-P_t^{2}}$, relation (\ref{eq3}) can be modified as:

\begin{equation}
|\frac{dP_t}{dt}| \leq \frac{2\Delta\hat{H}}{\hbar}\sqrt{P_t(1-P_t)}.
\label{eq34}
\end{equation}

Integrating (\ref{eq34}) for $P(0)=\cos^{2}\phi$, it is possible to write:

\begin{equation}
P_t \geq \cos^{2} (\frac {\Delta\hat{H}t}{\hbar}+\phi).
\label {eq35}
\end{equation}

In order to find the minimum time for the transition from the initial state $|a>$ to the target state $|c^{\perp}>$, orthogonal to $|c>$, the probability given by (\ref{eq33}) must be zero. Consequently,

\begin{equation}
t_{min}^{a \rightarrow c^{\perp}} = \inf {\{t \geq 0: P_t = 0\}},
\label {eq36}
\end{equation}
\noindent resulting:

\begin{equation}
t_{a \rightarrow c^{\perp}} \geq \frac{\hbar(\pi - 2\phi)}{2\Delta\hat{H}}.
\label{eq37}
\end{equation}

Therefore, for this case, the minimum time quantum control performance measure results:

\begin{equation}
\eta_{a \rightarrow c^{\perp}}^{H} = \frac{\hbar(\pi - 2\phi)}{2\Delta\hat{H}t_{CQS}}.
\label{eq38}
\end{equation}

\subsubsection{Transition between two general states}

Considering the states $|a>$ and $|b>$, with $|<b|a>|^{2} = \cos \phi$, and $\phi \in [0, \frac{\pi}{2}]$, if the dynamical evolution of the system is subjected to a temporal evolution operator $\hat{U}(t,t_0)$, with $\hat{U}(t,t_0) = e^{-{\frac{i\hat{H}(t-t_0)}{\hbar}}}$, for a time independent Hamiltonian $\hat{H}$, it is possible to study the state transition between $|a>$ and $|b>$ and calculate the minimum time transition:

\begin{equation}
t_{min}^{a \rightarrow b} = \inf {\{t \geq 0: P_t = |<b|\hat{U}(t,t_0)|a>|^{2}\}=1}.
\label {eq39}
\end{equation}

This calculation can be accomplished by following the steps proposed in \cite{5a}, considering that the evolution from $|a>$ to $|b>$ is equivalent to the evolution from $|a>$ to $|c^{\perp}>$, orthogonal to $|c>$, for any $|c>$ orthogonal to $|b>$.

As $|<b|a>|^{2}=\cos^{2}{\phi}$ with $\phi \in [0, \frac{\pi}{2}]$, the maximum value of $|<c|a>|^{2}$, when $|c>$ assumes all the states orthogonal to $|b>$, is  $\cos^{2}(\frac{\pi}{2}-\phi)$. Considering the sub-space of the kets $c=b^{\perp}$, orthogonal to $|b>$, expression (\ref{eq37}) can be applied. Finally, for the whole transition:

\begin{equation}
t_{a \rightarrow b} \geq \sup \{t_{a \rightarrow c^{\perp}}\} = \frac{\hbar \phi}{\Delta\hat{H}}.
\label{eq40}
\end{equation}

Therefore, for this case, the minimum time quantum control performance measure results:

\begin{equation}
\eta_{a \rightarrow b}^{H} = \frac{\hbar\phi}{\Delta\hat{H}t_{CQS}}.
\label{eq41}
\end{equation}

\section{Conclusions}

By using time-energy uncertainty relations, expressions for minimum time to quantum transitions were developed for three cases: state to orthogonal state transition; digital quantum computation, and general state transition.

These cases are useful for building quantum control and computation systems that need their efficiency evaluated. In order to perform this evaluation, a performance measure was defined and expressed for the three cases studied, considering minimum time quantum control systems.

It has been considered that the time evolution of the systems obeys transitions with time independent Hamiltonian. This hypothesis is compatible with quamtum computation and control processes that are supposed to occur in a very short period of time in order to avoid decoherence \cite{6b}.

\end{document}